\documentclass[twocolumn, aps, prl,superscriptaddress,showpacs,floatfix]{revtex4-1}

\usepackage{amsmath}
\usepackage{amssymb}
\usepackage{bm}
\usepackage{color}
\usepackage{graphicx}
\usepackage[latin1]{inputenc}
\usepackage[english]{babel}
\begin{document}


\title{Decoherence and mode-hopping in a magnetic tunnel junction-based spin-torque oscillator}
\author{P.~K.~Muduli}
\affiliation{Physics Department, University of Gothenburg, 41296 Gothenburg, Sweden}

\author{O.~G.~Heinonen}
\affiliation{Materials Science Division, Argonne National Laboratory, Lemont, IL 60439, USA}
\affiliation{Department of Physics and Astronomy, Northwestern University, 2145 Sheridan Rd., Evanston, IL 60208-3112}

\author{Johan~\AA kerman}
\affiliation{Physics Department, University of Gothenburg, 41296 Gothenburg, Sweden}
\affiliation{Materials Physics, School of ICT, KTH-Royal Institute of Technology,
Electrum 229, 164 40 Kista, Sweden}

\begin{abstract}
We discuss the coherence of magnetic oscillations in a magnetic tunnel junction-based spin-torque oscillator as a function of external field angle. Time-frequency analysis shows mode-hopping between distinct oscillator modes, which arises from linear and nonlinear couplings in the Landau-Lifshitz-Gilbert equation, analogous to
mode-hopping observed in semiconductor ring lasers. These couplings and therefore mode-hopping are minimized near the current threshold for antiparallel (AP) alignment of free layer with reference layer magnetization. Away from the AP alignment, mode-hopping limits oscillator coherence.
\end{abstract}

\pacs{85.75.-d, 75.78.-n, 72.25.-b,75.78.Cd}\maketitle


Magnetization precession at GHz frequencies can be
sustained in spin-valve (SV) and magnetic tunnel-junction
(MTJ) based spin-torque oscillators (STOs) by
directly transferring spin angular momentum from spin
polarized current~\cite{slonczewski1996jmmm,*berger1996prb,tsoi2000nt,*kiselev2003nt} to the free-layer magnetization
order parameter. A fundamental question is the stochastic phenomena that govern its
coherence time ($\tau_{\rm c}$). Theoretical studies~\cite{kim2008prl1,*kim2008prl,*slavin2009ieeem, silva2010ieeemag}
have investigated decoherence through thermal noise,
assuming that only a single mode is excited. Other theoretical works conclude that
only the lowest energy mode supports sustained oscillations~\cite{deAguiar2007prb}. Yet  experiments
clearly show the existence of multiple modes
 in SVs~\cite{kiselev2004prl,sankey2005prb,krivorotov2008prb,bonetti2010prl} and
 MTJs~\cite{nazarov2008jap,deac2008np,houssameddine2008apl,zheng2010prb,muduli2011prb},
and persistent mode-hopping~\cite{krivorotov2008prb,bonetti2010prl} between several modes. The impact on $\tau_{\rm c}$ of such mode-hopping
is largely unexplored, and theoretical investigations of its
origin are entirely lacking.

Here we present systematic experimental investigations
into mode-hopping and its impact on coherence
time, as functions of current
and applied field angle in MTJ-STOs. We derive equations showing how such systems are analogous
to semiconductor ring lasers (SRLs), and under driving can exhibit mode-hopping in the presence of stochastic noise.
Nonconservative torques in the Landau-Lifshitz equation couple individual modes.
The coupling has a minimum when the nonconservative torques cancel each other, explaining the experimentally
observed suppression of mode-hopping at
angles near (but not exactly at) antiparallel (AP)
alignment. Micromagnetic simulations of magnetization dynamics support this picture.
Finally, we show that although
mode-hopping is the limiting factor for STO coherence at
most angles, single-mode nonlinear spin-torque auto-oscillator
(NSTO) theory~\cite{kim2008prl1,*kim2008prl,*slavin2009ieeem} qualitatively holds when
dwell-time is sufficiently large compared with the oscillations' coherence time.

\begin{figure}[t!]
\includegraphics*[width=.45\textwidth]{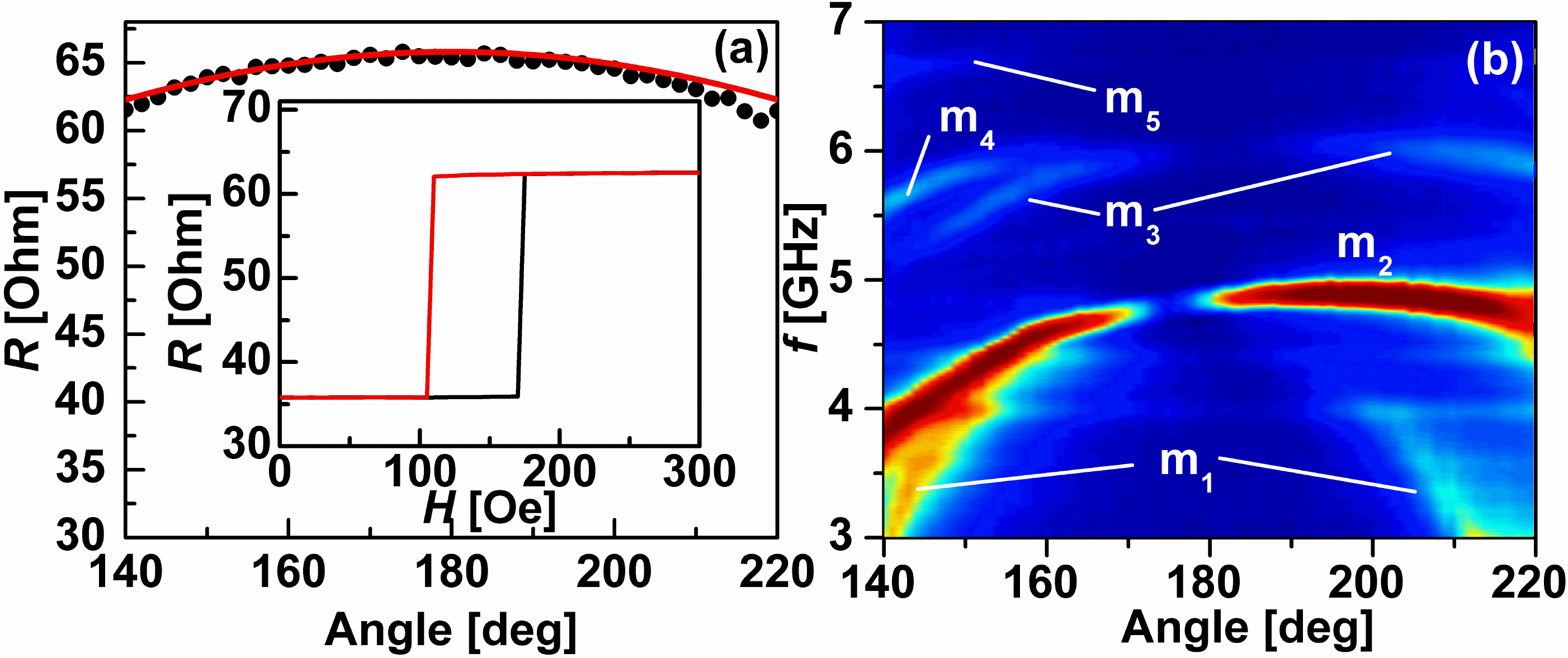}
\caption{(color online).(a) Experiment (circles) and calculated (solid line) resistance versus in-plane field angle $\varphi$ at $H$=450~Oe; $\varphi=180^\circ$ corresponds to AP alignment of FL and RL.
Inset: magnetoresistance loop measured at $\varphi=180^\circ$. (b)~Map of power (dB) vs. frequency ($f$) and $\varphi$ for $I$=8~mA and $H$=450~Oe.}\label{fig:fig1}
\end{figure}

The MTJ nanopillars used in this work are similar to those in Ref.~\onlinecite{muduli2011prb}. The layer structure consists of IrMn (5)/CoFe (2.1)/Ru (0.81)/CoFe (1)/CoFeB (1.5)/MgO (1)/CoFeB (3.5) (thickness in nm), where the bottom CoFe layer is the pinned layer (PL), the composite CoFe/CoFeB represents the RL, and the top CoFeB layer is the FL. We discuss results from a \emph{circular} device with approximate diameter 240~nm, resistance-area product 1.5~$\Omega~\mu$m$^{2}$, and tunneling magnetoresistance 75\%. The RL magnetization equilibrium direction is along the positive $\hat {x}$-direction, which is also $0^\circ$ of the applied field. We use the convention that a positive current flows from the FL to the RL.

\begin{figure}[t!]
\includegraphics*[width=.43\textwidth]{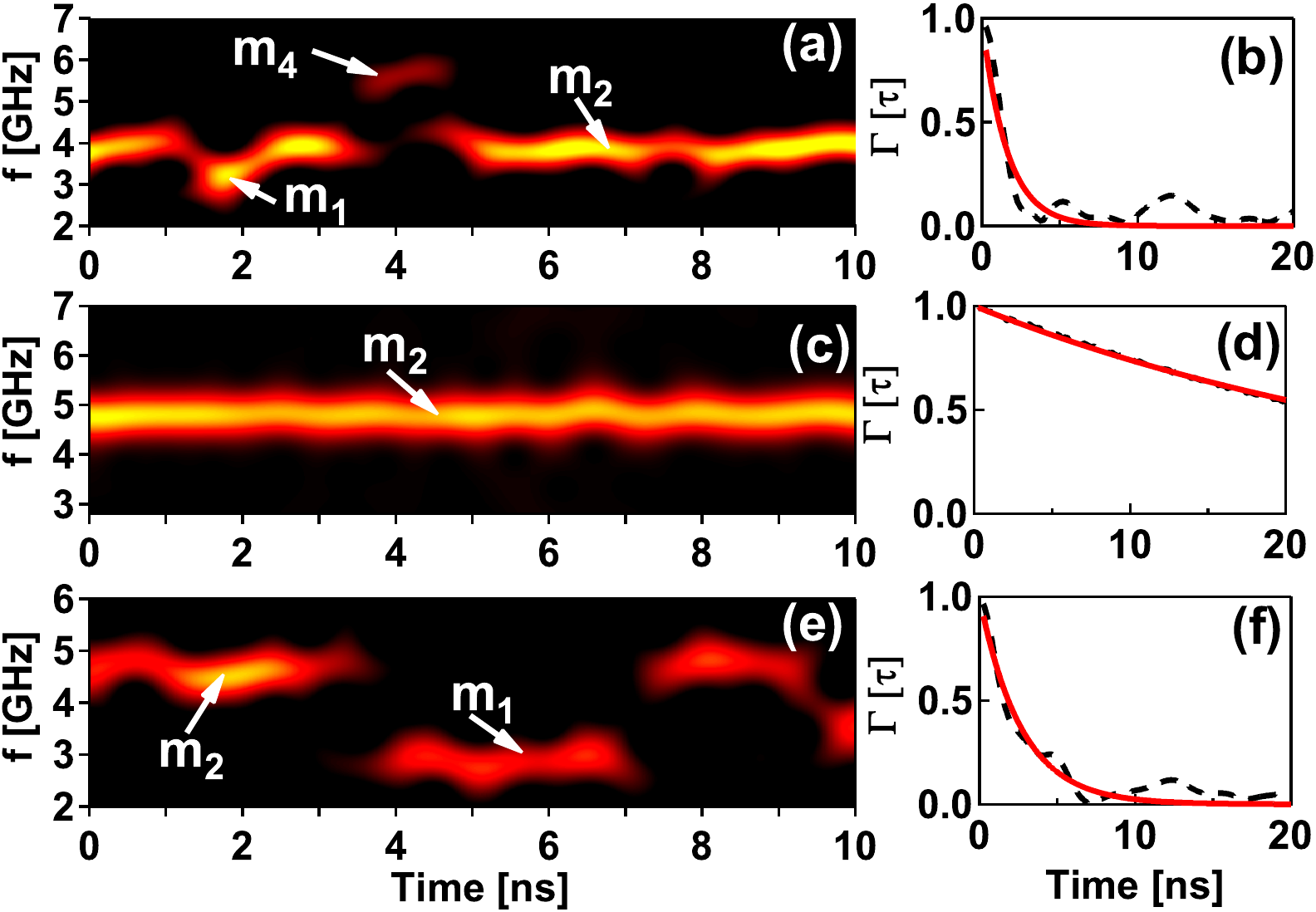}
\caption{(color online). Wigner distribution (first column), and normalized autocorrelation envelope (second column) at (a, b)~140$^\circ$, (c, d)~196$^\circ$, and (e, f)~$220^\circ$ for $I=$~8 mA and $H=$~450~Oe. Red lines are exponential fits to experimental autocorrelation envelope (dashed lines).}\label{fig:fig2}
\end{figure}

Figure~\ref{fig:fig1}(a) shows the resistance $R$ as a function of the in-plane angle $\varphi$ of applied field $H$.
Here we focus on magnetic excitations with $H=450$~Oe and $140^\circ\leq\varphi\leq220^\circ$, for which
the FL rotates {\em coherently} with the field [Fig.~\ref{fig:fig1}(a)]; with a positive current, FL modes are excited.
In general, several modes are found in frequency-domain measurements,
and we have identified five potential FL modes (m$_{i},i=1,\ldots,5$)~\cite{muduli2011prb}
from measurements at 8~mA. These modes' frequencies decrease when the field angle moves away
from 190$^\circ$, and are asymmetric about their maxima. Both the decrease and asymmetry are unexpected from a simple Kittel formula~\cite{muduli2011prb}, in which ferromagnetic
coupling between the FL and RL would result in mode frequencies symmetric about 180$^\circ$ and frequency minima at 180$^\circ$.

In order to quantitatively analyze the STO's time-frequency behavior, we collected 1,000~ns-long time-traces~\cite{suppinfo} of the
STO voltage signal at different field angles and bias currents. Figure~\ref{fig:fig2} shows the frequency vs. time obtained from Wigner transforms
of the traces at three representative angles,  $\varphi=140^\circ$, $196^\circ$, and $220^\circ$, and at 8~mA using a time (frequency) resolution of 1.25~ns (0.8~GHz).  We chose these angles because at
$140^\circ$ and $220^\circ$ the FL and RL are far from their AP configuration, but
the FL still rotates coherently with the external field, and at $196^\circ$ we observed longest coherence time of mode m$_2$,
because of the device's slight asymmetry [Fig.~\ref{fig:fig1}(b)] and because closer to
$180^\circ$, the observed m$_2$ coherence time is limited by frequency doubling~\cite{deac2008np,muduli2011jap}. The plots in
Fig.~\ref{fig:fig2} show that for 196$^\circ$, the STO mostly stays in mode m$_{2}$.  In contrast, frequent mode-hopping occurs between m$_{1}$ and m$_{2}$ at 220$^\circ$, and
between m$_{1}$, m$_{2}$, and m$_{4}$ at 140$^\circ$. We note that this mode-hopping behavior between distinct STO modes differs from the observations in Ref.~\onlinecite{houssameddine2009prl}, where the dominant mode's frequency fluctuated in time, on the scale $\sim0.01$~GHz. Detailed examination of time-frequency plots shows that mode-hopping occurs at all field angles and currents~\cite{suppinfo}, is completely random in time. The central mode m$_{2}$ is the most stable at all angles; spin torque preferentially excites mode m$_{2}$.

\begin{figure}[t!]
\includegraphics*[width=.35\textwidth]{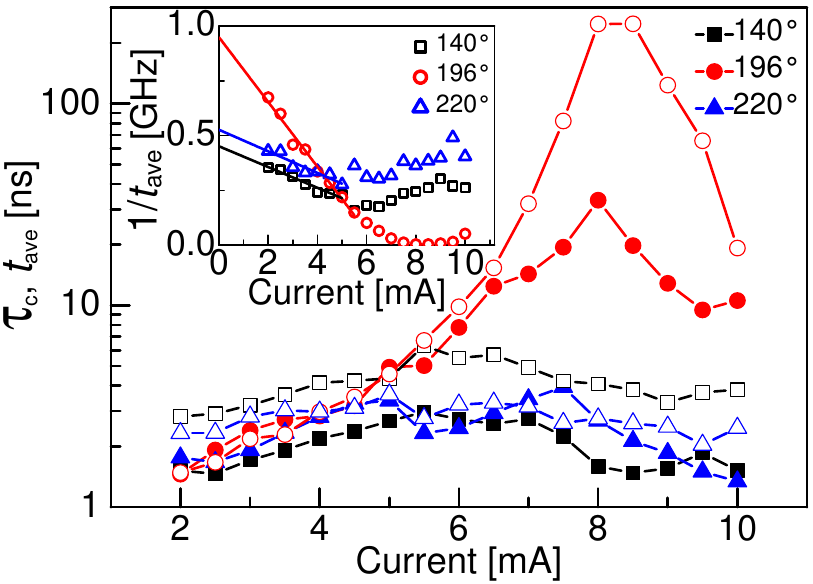}
\caption{(color online). Coherence time $\tau_{\rm c}$ (filled symbols) and average dwell-time $t_{\rm ave}$ (open symbols) vs. $I$ at 140$^\circ$ (black squares), 196$^\circ$ (red circles), and 220$^\circ$ (blue triangles). Inset: corresponding $1/t_{\rm ave}$ vs. $I$  showing a clear angular dependence of $1/t_{\rm ave}$ for $I\to0$.}\label{fig:coherence_time}
\end{figure}

The coherence time of mode m$_2$ can be obtained by autocorrelating time-traces. The right of Fig.~\ref{fig:fig2} shows the traces' normalized autocorrelation function, $\Gamma (\tau)$, filtered in a range of -300 MHz/+400 MHz around m$_2$. This filter width avoids overlapping other modes. The autocorrelation
functions decay exponentially in time, consistent with thermally activated stochastic processes leading to
decoherence~\cite{kramers1940phy,*vanthoff1884,*arrhenius1940zpc}. The corresponding decay time $\tau_{c}$, which is the coherence time for m$_2$, was obtained by fitting $\Gamma (\tau)$ to a function of the form $e^{-\tau/\tau_{c}}$.
We can compare $\tau_{\rm c}$ with the average dwell-time $t_{\rm ave}$ in mode m$_2$, obtained by analyzing the
instantaneous STO frequency~\cite{suppinfo}. Figure.~\ref{fig:coherence_time} shows the central experimental result: $\tau_{\rm c}$ and $t_{\rm ave}$ as functions of current at 140$^\circ$, 196$^\circ$, and 220$^\circ$. Both $\tau_{\rm c}$ and
$t_{\rm ave}$ depend on field angle and current. At 196$^\circ$, $t_{\rm ave}$ exceeds $\tau_{c}$ by over an order of magnitude for currents in a range about the threshold current. Here, mode-hopping is insignificant and does not limit coherence, the STO has a well-defined single mode, and we show below that NSTO theory applies here. For 140$^\circ$ and 220$^\circ$,  $\tau_{\rm c}$ and $t_{\rm ave}$ are approximately equal (within experimental uncertainty in $t_{\rm ave}$~\cite{suppinfo}). Here, mode-hopping limits coherence time; more extensive analysis shows that for $\varphi\alt165^\circ$ and $\varphi\agt205^\circ$, mode-hopping is the dominant decoherence process.

We now discuss the experimental results. The
magnetization dynamics can be described by the Landau-Lifshitz-Gilbert (LLG) equation, which we write as
\begin{equation}
\frac{d\hat m}{dt}  =  -{\hat m}\times \left[{\mathbf H}_{0}+{\mathbf h}_{\rm d}\right]
-{\alpha}{\hat m}\times
\left\{
{\hat m}\times \left[{\mathbf H}_0-\frac{a_{\rm J}}{\alpha}{\mathbf M}\right]+{\hat m}\times{\mathbf h}_{\rm d}
\right\},\nonumber
\label{eq:LLG_1}
\end{equation}
where ${\mathbf m}({\mathbf r})$ is the local magnetization direction of the FL and $\mathbf M$ that of the RL; ${\mathbf H}_0$ is the total {\em static}
effective field, including the out-of-plane spin-torque, ${\mathbf h}_{\rm d}$ is the {\em dynamic}
demagnetizing field arising from the oscillating magnetization density in FL, $a_{\rm J}$ is the in-plane effective field due to spin-torque, and $\alpha$ the dimensionless damping. First, micromagnetic simulations confirm that the
asymmetry in mode frequencies about $\varphi=180^\circ$ [Fig.~\ref{fig:fig1}(b)] can be caused by small ellipticity in the structure, with the exchange bias slightly misaligned
with the major axis, as the resonance modes are sensitive to the equilibrium magnetization details.
\begin{figure}[t!]
\includegraphics*[width=.49\textwidth]{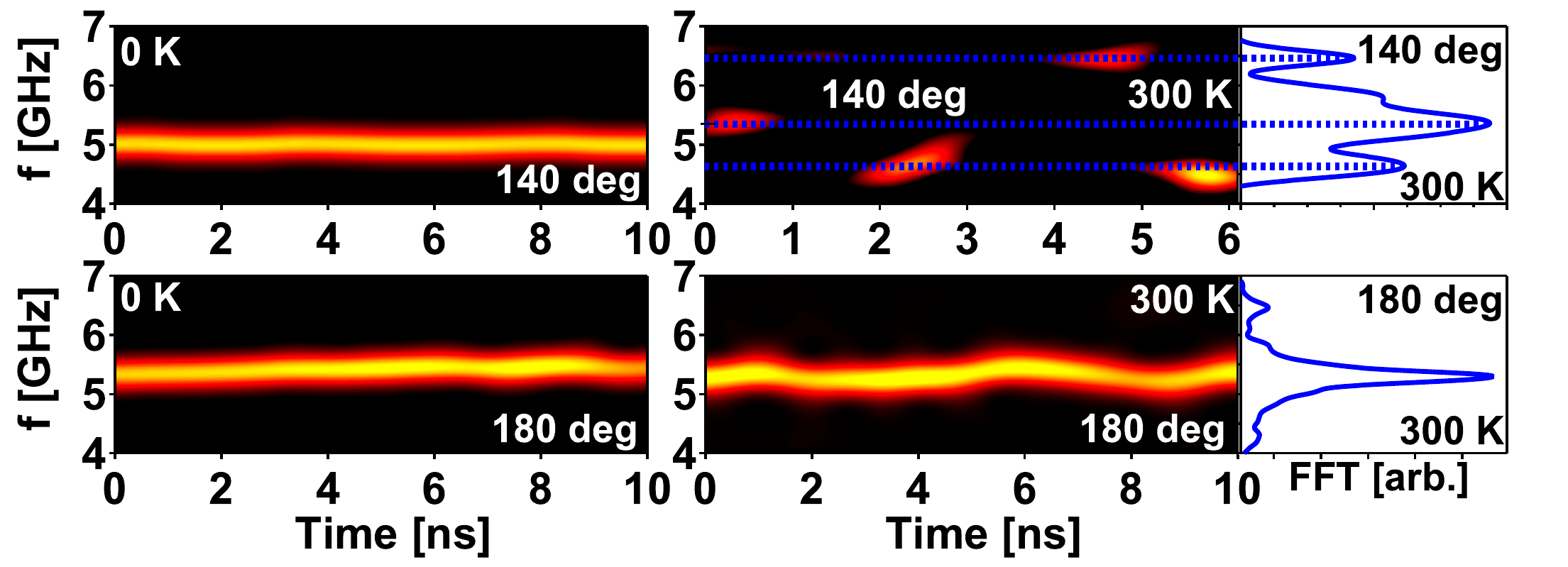}
\caption{(color online). Wigner transforms of micromagnetic time-traces of average FL magnetization at $T=0$~K (left)
and $T=300$~K (middle). The right represents the Fourier transform (FFT) of the 100~ns-long time-trace at 300~K. The top row is for 140$^\circ$; the bottom row for 180$^\circ$. Bias current was 8~mA.}\label{fig:micromag_trace}
\end{figure}
The magnetoresistance, however, remains nearly symmetric about $\varphi=180^\circ$, as it measures the {\em average} angle
between RL and FL magnetization directions.
Now consider a system with the equilibrium FL magnetization along ${\mathbf H}_0$, which is at an angle $\phi$ to the applied field. The linearized (about static equilibrium) LLG equation can be recast as an
eigenvalue equation ${\cal L}({\mathbf r},{\mathbf r}';\omega){\mathbf m}({\mathbf r}')=0$. At zero damping and current, all torques in the LLG equation are
conservative, and the eigenvalues are
real and undamped, but finite damping or current gives rise to nonconservative torques, and eigenvalues become complex.
We now outline how this leads to mode-coupling~\cite{suppinfo}.
We restrict the discussion to two modes, and
write the magnetization texture as
$\hat m({\mathbf r},t)=c_1(t)\hat m_1({\mathbf r}){\rm e}^{-i\omega_1 t}+c_2(t)\hat m_2({\mathbf r}){\rm e}^{-i\omega_2 t}$,
where $\hat m_i({\mathbf r})$ is an eigenmode of the {\em linearized} LLG with eigenfrequency $\omega_i$;
the coefficients $c_i$ carry the slow time dependence (on observational timescales $\tau_{\rm obs}$) of the magnetization texture. We insert this into the full nonlinear LLG,
project onto $\hat m_i$, and average over fast timescales, before expanding the nonlinear projections in $c_i$. We write the result as
\begin{widetext}
\begin{eqnarray}
\dot{c}_1
&=& -i\left[\omega_1\eta_{1,1}|c_1|^2+\omega_2\eta_{1,2}|c_2|^2\right]c_1-\Gamma_G\left[1+P_{1,1}\omega_1|c_1|^2
+P_{1,2}\omega_2|c_2|^2\right]c_1
+\sigma_0I\left[1-Q_{1,1}\omega_1|c_1|^2-Q_{1,2}\omega_2|c_2|^2\right]c_1+R_{1,2}c_2\label{eq:LLG_reduce_1_1}\nonumber\\
\dot{c}_2
&=& -i\left[\omega_1\eta_{2,1}|c_1|^2+\omega_2\eta_{2,2}|c_2|^2\right]c_2-\Gamma_G\left[1+P_{2,1}\omega_1|c_1|^2
+P_{2,2}\omega_2|c_2|^2\right]c_2
+\sigma_0 I \left[1-Q_{2,1}\omega_1|c_1|^2-Q_{2,2}\omega_2|c_2|^2\right]c_2+R_{2,1}c_1,\nonumber
\label{eq:LLG_reduce_1_2}
\end{eqnarray}
\end{widetext}

where $\eta_{i,j}$, $P_{i,j}$, and $Q_{i,j}$ are real; $R_{i,j}$ is complex. The factors $\eta_{i,j}$ are nonlinear
frequency shifts, $P_{i,j}$ the nonlinear positive damping, $Q_{i,j}$ the nonlinear negative damping, and $\sigma_0$ the usual spin-torque coefficient defined in Ref.~\onlinecite{slavin2009ieeem}. These equations are generalizations
of the equation for the single-mode NSTO~\cite{slavin2009ieeem}.
We note the linear ``backscattering'' term $R_{1,2}$ ($R_{2,1})$ in these equations. 
This term does not appear in normal spin-wave expansions
of Hamiltonians or equations of motion as it violates energy conservation on short timescales. Here, however, we
consider conditions close to threshold: energy is approximately conserved
on long time-scales, and we enforce the constraint $\omega_1|c_1|^2+\omega_2|c_2|^2=p$,where $p$ is a constant.
On general grounds, we can set $Q_{1,2}=Q_{2,1}=Q_o$,
$P_{1,2}=P_{2,1}=P_o$, and $R_{1,2}=R_{2,1}=u+iv$. We introduce new variables $Q_i{\rm e}^{i\phi_i}=\sqrt{\omega_i}c_i$ and $\psi=\phi_2-\phi_1$ and use
the constraint $Q_1\dot{Q}_1+Q_2\dot{Q}_2=0$. The equations for slow time evolution can then be recast as
a two-dimensional $Z_2$-invariant dynamical system, analogous to that describing SRLs~\cite{vanderSande,beri2008prl}. Close to
threshold, these
equations are known~\cite{beri2008prl}
to have two stable solutions (either of the two modes) close to a
homoclinic bifurcation. In the presence of a stochastic
field originating from, {\it e.g.\/} contact with a thermal bath, competition between linear and nonlinear couplings can lead to mode-hopping~\cite{ohtsu1986IEEE_JQE,vanderSande,beri2008prl} between the two stable solutions.
Nonconservative torques lead to linear and nonlinear couplings between modes, as expressed by the terms in $Q_{i,j}$,
$P_{i,j}$, and $R_{i,j}$, $i\not= j$.
As the bias current increases from zero, the local nonconservative field is ${\mathbf m}\times \left[{\mathbf H}_0-\frac{a_{\rm J}}{\alpha}
{\mathbf M}\right]$; this field also dominates the damping (imaginary parts of eigenvalues). For arbitrary field angles,
a current exists---the threshold current--at which the energy dissipation rate equals the rate at
which energy is pumped into the system by the spin
torque. At this current, energy conservation $Q_1^2+Q_2^2=0$ is strict, but the couplings between modes do not vanish,
and the system exhibits mode-hopping: the system's total energy is conserved, but energy is transferred back and
forth between the modes. However, at 180$^\circ$, ${\mathbf H}_0$ and ${\mathbf M}$ are collinear, and at threshold energy dissipation equals
energy pumping {\em and} local dissipative torques vanish. Ignoring the nonlocal dissipative torques, this
means that {\em all} terms multiplying $\Gamma_G$, $\sigma_0 I$, as well as $R_{i,j}$, vanish. Consequently, mode-hopping is minimized. The system's total energy is conserved, and each mode's energy is also individually conserved.

Figure.~\ref{fig:micromag_trace} shows Wigner transforms of magnetization time-traces obtained from micromagnetic modeling for a system similar to the experiment~\cite{suppinfo}. At a finite temperature of $T=300$ K and $I=8$ mA, mode-hopping occurs at 140$^\circ$, while at 180$^\circ$ precession is coherent at the m$_2$ mode---this current is just above the threshold for this field angle. Yet at $T=0$~K, no mode-hopping occurs. Apart from demonstrating that mode-hopping exists within the micromagnetic model, the modeling shows that (a) mode-hopping is induced by thermal fluctuations, and (b) the angle-dependence of
mode-hopping agrees qualitatively with our analysis above.

\begin{figure}[t!]
\includegraphics*[width=.45\textwidth]{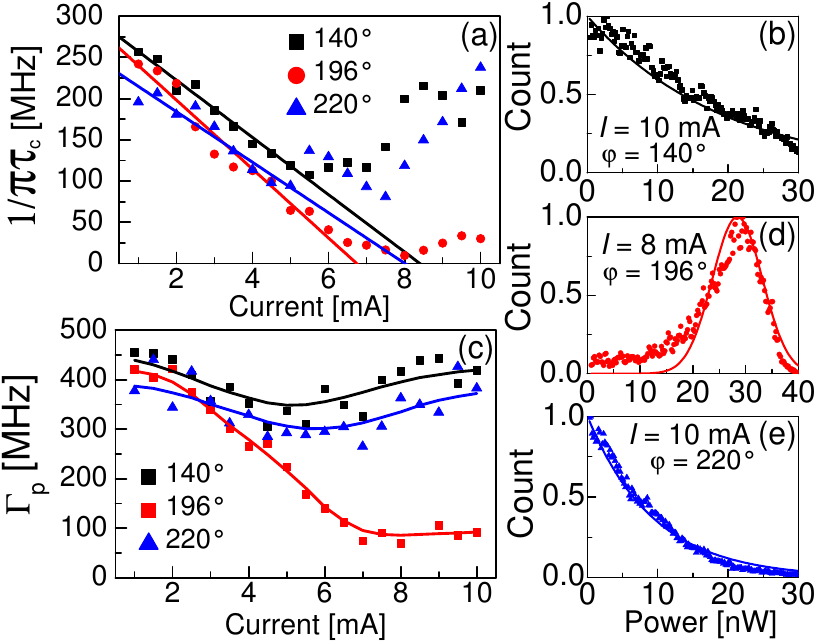}
\caption{(color online). (a) $1/(\pi\tau_{\rm c})$ vs. current at 140$^\circ$ (black squares), 196$^\circ$ (red circles), and 220$^\circ$ (blue triangles). Solid lines are fits to the subthreshold region; Intercepts with the horizontal axis indicate threshold currents. (c) Power restoration rate $\Gamma_{\rm p}$ vs. current at $140^\circ$ (black squares),
$196^\circ$ (red circles), and $220^\circ$ (blue triangles). Solid lines are visual guides. Right: Normalized power distribution function and fits at (b)~$140^\circ$, (d)~$196^\circ$, and (e)~$220^\circ$.
}\label{fig:fig4}
\end{figure}

If the mode resident time $t_{\rm ave}$ is long enough, one can analyze the system in terms of single-mode
NSTO theory~\cite{slavin2009ieeem}. The experimental coherence time $\tau_{\rm c}$ can then
be compared with that from NSTO theory. Under the assumption that $\tau_{\rm c}\propto 1/\Delta f$, with
$\Delta f$ the linewidth,
we plot $1/(\pi\tau_{\rm c})$ against bias current and fit $1/(\pi\tau_{c})$ to the NSTO expression~\cite{slavin2009ieeem} for subthreshold linewidth,
$\Delta f=\Gamma_{\rm G}(1-I/I_{\rm th})$, where $\Gamma_{\rm G}$ is the natural FMR linewidth,
and $I_{\rm th}$ the threshold current [Fig.~\ref{fig:fig4}(a)]. For $196^\circ$ we obtain $\Gamma_{\rm G}\approx300$~MHz and $I_{\rm th}=6.4$~mA. This $\Gamma_{\rm G}$ is comparable to previous reports~\cite{georges2009prb} and agrees well with an estimate using these material parameters for FL: saturation magnetization $M_{0}=1000$~emu/cm$^{3}$, Gilbert damping parameter $\alpha_{\rm G}=0.01$. Above $I_{\rm th}=6.4$~mA, $1/(\pi\tau_{\rm c})$ increases with current, qualitative agreeing with the NSTO theory prediction. Thus the linewidth from NSTO theory qualitative describes $1/(\pi\tau_{\rm c})$ at $196^\circ$ well~\cite{suppinfo}, consistent with $t_{\rm ave}$ being large enough that the STO is well described as a single-mode
oscillator with decoherence caused by single-mode thermal fluctuations. However for $140^\circ$ and $220^\circ$ the coherence is now limited by mode-hopping, and we show that the strong increase of $1/(\pi\tau_{\rm c})$ above 8~mA is primarily due to increased mode-hopping, not due to increased nonlinearity in NSTO theory.

Figure~\ref{fig:fig4} (c) shows the measured power restoration rate $\Gamma_{\rm p}$~\cite{bianchini2010apl}.
According to NSTO theory, $\Gamma_{\rm p}$ vanishes as $\Gamma_{\rm G}(I/I_{\rm th}-1)$ near the threshold.
Again, for $196^\circ$ the agreement between the measured power restoration rate and the NSTO theory prediction is reasonably good. But for $140^\circ$ and $220^\circ$ the power restoration rates, like the linewidths, have much larger minimum values, consistent with a limiting decoherence process other than thermal fluctuations about m$_2$. We note that even for $196^\circ$, $\Gamma_{\rm p}$ does not vanish at $I=I_{\rm th}$, indicating either nonzero mode-hopping or the presence of additional sources of noise. Further inconsistencies between NSTO
theory and experimental data is provided by the power distribution functions shown on the right of Fig.~\ref{fig:fig4}. The current values are chosen such that $I/I_{\rm th}=1.25$ in all cases. According to NSTO theory,
the power distribution function ${\cal P}(p)$ has the form of a Gaussian~\cite{tiberkevich2007apl} $\exp[-(p-p_{0})^{2}/2\Delta p^{2}]$ for $I>I_{\rm th}$, where $p_{0}$ denotes stationary power and $\Delta p$ the power fluctuations. This agrees well with the measured power distribution function  near $\varphi=196^\circ$, shown by the solid line in Fig.~\ref{fig:fig4}(d), with the relative power fluctuations $\Delta p/p_{0}$
in the $0.2$--$0.4$ range, similar to that obtained in Ref.~\cite{nagasawa2011ape}. In contrast, the power distributions are exponential at 140$^\circ$ and 220$^\circ$, consistent with below-threshold conditions.
We conclude that the threshold currents for these angles extracted from the linewidth measurements using NSTO theory are too small.
Consequently, the increase in measured linewidth $1/(\pi\tau_{\rm c})$ above 8~mA at $140^\circ$ and $220^\circ$
\emph{cannot} be attributed to an increase in nonlinearity,
unlike Ref.~\onlinecite{georges2009prb}. This is further supported by the fact that the experimental nonlinear frequency shift did not show any strong angular dependence~\cite{suppinfo}.

One can qualitatively understand how moderate mode-hopping changes NSTO theory. First, it leads to effectively increased power dissipation from mode m$_2$, since hopping decreases the power in this mode.
Second, for short excursions to other modes, the oscillator phase is random as it returns to m$_2$, and is not correlated with the phase
before the excursion. This means mode-hopping adds extra phase noise to oscillators. Within this simple picture, we write the expression for the subthreshold linewidth as $\Delta f\approx \Gamma_{\rm G}-\sigma_0 I+\frac{a}{t_{\rm ave}}$; $a$ is a dimensionless constant of order unity. 
Since $a/t_{\rm ave}$ depends on the angle for $I\to0$ (see Fig.~\ref{fig:coherence_time}, inset), this explains the obtained apparent angular dependence of the zero-current linewidth $\Gamma_{\rm G}$ [Fig.~\ref{fig:fig4}(a)]. Furthermore, the extrapolated
threshold current value shifts to a larger value, the shift being larger for weaker current-dependence of $t_{\rm ave}$, consistent with a larger threshold current for $140^\circ$ and $220^\circ$ than that extrapolated
using NSTO theory.

In summary, we have observed and analyzed multi-mode excitations and mode-hopping in an MgO-based spin-torque oscillator (STO).
Mode-hopping occurred at all angles and currents, contrary to two-mode theory~\cite{deAguiar2007prb} and expectations~\cite{slavin2009ieeem}. Insofar as the basic physics of mode-hopping appears to be intrinsic to the
LLG equations governing the dynamics of the oscillator nano-sized systems with single-domain equilibrium magnetization, efforts to reduce decoherence may need to focus
on better understanding the energy barrier separating stable modes.

We acknowledge G.~Finocchio, S.~Bonetti, and Randy K.~Dumas for useful discussions. Support from the Swedish Foundation for Strategic Research (SSF),
the Swedish Research Council (VR), the G\"{o}ran Gustafsson Foundation and
the Knut and Alice Wallenberg Foundation are gratefully acknowledged. J.~\AA. is a Royal Swedish Academy of Sciences Research Fellow supported by a grant from the Knut and Alice Wallenberg Foundation. Argonne National Laboratory is operated under Contract No. DE-AC02-06CH11357 by UChicago Argonne, LLC.

\end{document}